# Polaronic metal state at the LaAlO$_3$/SrTiO$_3$ interface

C. Cancellieri[1,2,*], A.S. Mishchenko[3,*], U. Aschauer[4], A. Filippetti[5], C. Faber[4], O.S. Barišić[6], V.A. Rogalev[1], T. Schmitt[1], N. Nagaosa[3] and V.N. Strocov[1,*]

[1] Swiss Light Source, Paul Scherrer Institute, CH-5232 Villigen-PSI, Switzerland

[2] EMPA, Ueberlandstrasse 129, 8600 Duebendorf, Switzerland

[3] RIKEN Center for Emergent Matter Science, 2-1 Hirosawa, Wako, Saitama 351-0198, Japan

[4] Materials Theory, ETH Zurich, Wolfgang-Pauli-Strasse 27, CH-8093 Zürich, Switzerland

[5] CNR-IOM, Istituto Officina dei Materiali, Cittadella Universitaria, Monserrato (CA) 09042-I, Italy

[6] Institute of Physics, Bijenička 46, 10000 Zagreb, Croatia

* These authors have contributed equally to this work. Correspondence should be addressed to V.N.S. (email: vladimir.strocov@psi.ch)

**Interplay of spin, charge, orbital and lattice degrees of freedom in oxide heterostructures results in a plethora of fascinating properties, which can be exploited in new generations of electronic devices with enhanced functionalities. The paradigm example is the interface between the two band insulators LaAlO$_3$ and SrTiO$_3$ (LAO/STO) that hosts two-dimensional electron system (2DES). Apart from the mobile charge carriers, this system exhibits a range of intriguing properties such as field effect, superconductivity and ferromagnetism, whose fundamental origins are still debated. Here, we use soft-X-ray angle-resolved photoelectron spectroscopy to penetrate through the LAO overlayer and access charge carriers at the buried interface. The experimental spectral function directly identifies the interface charge carriers as large polarons, emerging from coupling of charge and lattice degrees of freedom, and involving two phonons of different energy and thermal activity. This phenomenon fundamentally limits the carrier mobility and explains its puzzling drop at high temperatures**.

Coupling of the electron and lattice degrees of freedom in solids through electron-phonon interaction (EPI) is a key concept in electron transport and many other phenomena of condensed matter physics. An electron moving in the lattice can displace atoms from their equilibrium positions in response to the EPI. Such an electron (or hole) dragging behind a local lattice distortion - or phonon "cloud" - forms a composite charge carrier known as polaron[1,2]. The increased effective mass $m^*$ of this quasiparticle fundamentally limits its mobility $\mu \propto 1/m^*$ beyond the incoherent scattering processes. The polarons are key players in many technological devices, a widespread example of which are high electron mobility transistors (HEMTs) utilized in high-frequency devices such as mobile phones. In typical HEMTs, a donor layer of n-doped AlGaAs injects electrons into the channel layer of intrinsic GaAs where, escaping scattering on the dopant impurities, the electrons are limited in their mobility only by the polaronic coupling enhanced by spatial confinement in the GaAs quantum well (QW)[3].

Angle-resolved photoelectron spectroscopy (ARPES) is a unique method to measure the single particle spectral function $A(\omega,\mathbf{k})$ in crystalline solids resolved in electron energy $\omega$ and momentum $\mathbf{k}$. Containing all many-body (electron-electron, electron-phonon, etc.) interactions, $A(\omega,\mathbf{k})$ reveals the formation of polarons by a characteristic peak-dip-hump (PDH) lineshape, where the sharp peak corresponds to a quasiparticle (QP) and the broad hump, extending to higher binding energies, corresponds to the cloud of entangled phonons with frequencies $\omega_0$[1,2]. However, the extreme surface sensitivity of conventional ARPES with photon energies $h\nu$ below ~100 eV sets the buried interfaces out of its reach. The crucial feature of our experiment is the use of soft-X-ray ARPES (SX-ARPES) operating in the $h\nu$ range of hundreds eV (for a recent review see Ref. 4). The longer photoelectron mean free path enables SX-ARPES to penetrate through the top layers and access $A(\omega,\mathbf{k})$ at buried interfaces.

Complex oxide interfaces are presently at the forefront of fundamental research in view of their enhanced functionalities achieved by exploiting electron correlations[5,6]. The 2DES in LAO/STO[7] is confined within a narrow region of a few nanometres on the STO side[8,9,10], where the mobile electrons populate the $t_{2g}$-derived $d_{xy}$-, $d_{xz}$- and $d_{yz}$-states of Ti ions acquiring reduced valence compared to the bulk $Ti^{4+}$. Confinement in the interface QW further splits these states into a ladder of subbands[8,11,12,13,14]. This complex energy structure based on the correlated $3d$ orbitals, very different from conventional semiconductor heterostructures described as free particles embedded in the mean-field potential, is the source of a rich and non-trivial phenomenology. Here, high 2DES mobility $\mu_{2DES}$ typical of uncorrelated electron systems co-exists with superconductivity[5,6], ferromagnetism[15], large magnetoresistance[17] and other phenomena typical of localized correlated electrons. Other intriguing puzzles in this intricate physics are why $\mu_{2DES}$ measured in transport falls short of estimates based on mean-field theories, and what causes the dramatic drop of $\mu_{2DES}$ with increase of temperature[18].

Here, we directly access the nature of the LAO/STO interface carriers through their $A(\omega,\mathbf{k})$ measured by SX-ARPES at ultrahigh energy resolution. We discover that the LAO/STO interface forms a *polaronic metal state* involving at least two active phonons. Whereas polaronic coupling to hard LO3 phonons fundamentally limits $\mu_{2DES}$ at low temperatures, coupling to soft TO1 phonons with increasing temperature provides the microscopic mechanism of $\mu_{2DES}$ drop observed in transport.

**Spectroscopic signatures of the polaronic metal state**

Our LAO/STO(001) samples with an LAO overlayer thickness of ~18 Å corresponding to 5 unit cells (u.c.) were grown using Pulsed Layer Deposition (PLD), and subsequently annealed in oxygen atmosphere to minimize the concentration of oxygen vacancies ($V_O$s) and the related extrinsic charge carriers[19]. SX-ARPES with its crucial advantage of enhanced probing depth is ideally suited to study this buried system where the 2DES only develops with the LAO layer thickness above 3 u.c.[20] The extremely small 2DES signal, however, has to be boosted using resonant photoemission[21] locked to the interface Ti ions[14,22,23]. For details of the sample growth and SX-ARPES experiment, see Methods.

Our low-temperature experimental dataset in Fig. 1 was measured at 12K using $s$-polarized X-rays (the parallel $p$-polarization data are given in the Supplementary 1). The resonance map of (angle-integrated) photoemission intensity, Fig 1a, was recorded under variation of $hv$ across the Ti $2p$ absorption edge around 460 eV. We identify there the 2DES signal at the Fermi level $E_F$ blowing up near the two $Ti^{3+}$ $L_3$- and $L_2$-resonances and vanishing everywhere else (we notice that strong suppression of the in-gap states around -1.2 eV, related to the $V_{OS}$ [14,22,23], confirms the prevalence of the intrinsic interface charge carriers). Tuning $hv$ onto the stronger $L_3$-resonance produces the Fermi surface (FS) map in Fig. 1b where, by comparison with the superimposed theoretical FS contours, we recognize the manifold of merged circular $d_{xy}$-derived FS sheets and the elliptical $d_{yz}$-sheet extending in the $k_x$-direction[14,23].

The ARPES images measured along the ΓX ($k_y$=0) line of the square two-dimensional Brillouin zone (BZ) at the $L_3$- and $L_2$-resonances are shown in Fig. 2a and b, respectively. With a high energy resolution of 40 meV, these images resolve individual interface bands. The use of $s$-polarization selects the $d_{xy}$- and $d_{yz}$-derived states, which are antisymmetric relative to the ΓX line[14] (although the selection rules are slightly relaxed by remnant structural distortions). By comparison with the overlaid $E(\mathbf{k})$ dispersions calculated with pSIC DFT (see Methods) one can recognize the lower $d_{yz}$-band with its flat dispersion. Already at this point, we note signs of a quasiparticle interaction, which reduces the band dispersion compared to the overlaid DFT prediction as characterized by an effective mass ratio of $m^*/m_0 \sim 2.5$. The $d_{xy}$-bands are not visible due to vanishing matrix elements, but the lowest $d_{xy}$-band appears as two bright spots where it hybridizes with the $d_{yz}$-band.

The most striking visual aspect of the experimental $E(\mathbf{k})$ is, however, the two vertical waterfalls extending down from these high-intensity $d_{xy}$ spots. In Fig. 2c,d we show the energy distribution curves (EDCs) – i.e. ARPES intensity as a function of binding energy for a given $\mathbf{k}$ – extracted from the images in Fig 2a and b, respectively. These EDCs reveal, remarkably, a pronounced PDH structure of $A(\omega,\mathbf{k})$, where the peak reflects the QP and the hump at ~118 meV below the peak its coupling to bosonic modes (such as magnon, plasmon, phonon) whose nature will be identified later on.

We note in passing that the EDC representing the whole $d_{xy}$-band, Fig. 2c, exhibits a smaller but broader QP peak in comparison to its $d_{yz}$-counterpart in Fig. 2d. Their nearly equal integral QP weight indicates that, non-trivially, the bosonic coupling is quite insensitive to different spatial distribution of the $d_{xy}$- and $d_{yz}$-states[8] The larger broadening of $d_{xy}$-EDC can reflect larger defect scattering, because the lowest $d_{xy}$-state is localized closest to the interface where the concentration of defects generated by the non-equilibrium PLD growth is maximal, whereas the $d_{yz}$-state extends deeper into the defect-free STO bulk. This observation is consistent with the recent Shubnikov–de Haas experiments[15] which have found smaller $d_{xy}$-mobility compared to $d_{xz/yz}$-one.

We will now address the nature of the involved bosonic modes. Considering the extremely small ferromagnetic response of the LAO/STO interface[15], the magnons can safely be ruled out. Plasmons can also be excluded since the energy of the hump does not depend on the interfacial carrier concentration $n_s$ varied via manipulation of $V_{OS}$ (see Supplementary 2) while the plasma frequency $\omega_p$ is proportional to $\sqrt{n_s}$. These bosonic modes assign therefore to phonons coupling to electron excitations and forming polarons[24,25,26]. The hump apex, located at ~118 meV below the QP peak, identifies the main coupling phonon frequency $\omega_0'$. The 2DES at the LAO/STO interface realizes therefore a *polaronic metal state*.

To identify the phonon modes forming the observed polaron, we used DFT to calculate the phonon dispersions (see Methods) for cubic bulk STO at different electron doping concentrations $n_V$; the results are shown in Fig. 3a. Although these initial calculations are for cubic STO, we will address shifts of the phonon frequencies at the cubic to tetragonal phase transition when discussing the temperature dependence. The $\frac{\partial \omega_0}{\partial n_V}$ response of the calculated dispersions quantifies the EPI strength as a function of phonon mode and **q**. With the actual electron density distribution[8,9,10], our $n_s$ measured by the Luttinger area of the experimental FS roughly corresponds to $n_V = 0.12$ electrons/u.c. Since the carriers reside on the STO side of the LAO/STO interface, we expect these results to be relevant for the interface as well. The polaron can be associated with the hard longitudinal optical phonon LO3, which is the only mode available in this high energy range and, moreover, has the largest coupling constant λ among all LO modes[27,28] as observed by Raman[29] and neutron spectroscopy[30]. At finite **q** vectors, the LO3 mode represents a breathing distortion of the octahedral cage around a Ti site, Fig. 3b, which is typical of polaron formation driven by the (Holstein-type) *short-range EPI*, barely sensitive to electron screening in a metallic system. This character of EPI is confirmed by the increase of $\frac{\partial \omega_0'}{\partial n_V}$ with increase of **q** away from the Γ-point. Recent optical studies on bulk STO[31] have not only confirmed the involvement of the LO3 phonon in EPI but also found the corresponding effective mass renormalization $m^*/m_0$ ~3.0 close to our value. This mode has also been observed by ARPES on (doped) bare STO(001)[32,33] with the strength of the polaronic structure depending on $n_s$. We note that, our theoretical LO3 energy of ~100 meV perfectly matches that found experimentally for the bare STO bulk and surface, but it differs from our $\omega_0'$ ~ 118 meV measurement at the LAO/STO interface. To gain some insight on this discrepancy, we investigated the possible role of strain by performing additional calculations with the STO in-plane lattice constants constrained to that of LAO. However, only a minor frequency shift to 102 meV was found. This implies that other interfacial factors such as the electric field, phonon coupling across the interface, propagation of the LAO distortion into STO[34], structural changes due to electrostatic doping[35] or weaker coupling to additional phonon modes could also play a role. We also note that the EPI at LAO/STO is enhanced by the tight 2D electron confinement in the interface QW[2,3], which was recently evidenced by huge oscillations of thermopower as a consequence of large phonon drag[36].

Next, we perform a theoretical analysis of the PDH structure to estimate the strength of the EPI governing the polaron formation. With the total $A(\omega,\mathbf{k})$ spectral weight normalized to unity, we define $Z_0$ as the integral weight of the QP part. The observed $A(\omega,\mathbf{k})$ of a polaron with momentum $\mathbf{k}$ can be expressed as a sum of two terms[1,2]

$$A(\omega,\mathbf{k}) = Z_0 \delta[\omega - E(\mathbf{k})] + A^H(\omega,\mathbf{k}) \quad (1)$$

The first term represents the sharp QP peak with the dispersion $E(\mathbf{k})$ and effective mass $m^*$ resulting from the renormalization of the non-interacting single-particle band with dispersion $\varepsilon(\mathbf{k})$ and mass $m^0$. The second term arises due to phonons coupling to the excited photohole. The EPI has a twofold effect on $A(\omega,\mathbf{k})$: it reduces $Z_0$ below unity, and builds up $A^H(\omega,\mathbf{k})$, with its dispersion following $\varepsilon(\mathbf{k})$, as a Frank-Condon series of phonon peak replicas at energy separations $\omega_n = n\omega_0$ from the QP peak, where $\omega_0$ is the phonon frequency and $n$ indexes the replicas.

For our analysis we used the angle-integrated EDC at the $L_3$-resonance (Fig. 1e) which is dominated by the $d_{xy}$-intensity and thus insensitive to the dispersion effects in the $d_{yz}$-band (certain admixture of the latter is not important because its dispersion range is much smaller than our experimental resolution of 40 meV and, moreover, the $d_{xy}$ and $d_{yz}$ bands have the same peak-to-hump relative weight). Gaussian fitting of its QP peak yields $Z_0$ ~ 0.4. For the short-range EPI, forming the LO3 polaron, exact diagrammatic quantum Monte Carlo (QMC) calculations[37] have shown that the relation $m^*/m_0$ ~ $1/Z_0$ holds with high accuracy. With the determined $Z_0$, this yields $m^*/m_0 = 2.5$, which coincides with the value extracted above from the band dispersions, leaving not much room for electron correlations to contribute to the band renormalization. The latter is confirmed by theoretical analysis of tunneling experiments[38] where the local electron correlations were described with the Hubbard parameter $U$. According to Ref. 39, already at $U = 4t$ ($t$ is the hopping integral) the density of states significantly renormalizes and forms a pseudogap at $E_F$. The absence of the latter in the tunneling spectra[38] indicates $U < 4t$. Then the renormalization expected from the correlation effects $m*/m_0 = [1-(U/U_c)^2]^{-1}$ ($U_c=8t$ is the critical Mott value) is less than 1.3, which is much smaller than our experimental $m*/m_0$. Hence, the mass renormalization at the LAO/STO interface is dominated by EPI.

Furthermore, the coupling strength λ for our short-range EPI can be estimated as $λ ~ 1 - Z_0$, which yields $λ ≈ 0.6$. This value is well below the limit $λ ~ 0.9$ separating the weak- and strong-coupling regimes, which implies that we observe *large polarons,* where the lattice distortion extends over several unit cells[1,2,39]. This is perfectly consistent with the clear dispersion of the hump, which tracks the non-interacting $ε(\mathbf{k})$ dispersion of the $d_{yz}$-band in Fig. 1d. Importantly, the small λ also excludes that self-trapping of small polarons[41,42] can be responsible for the discrepancy between the observed mobile charge and the 0.5 electrons/u.c. required for full compensation of the polar field in the LAO overlayer[5,6,42]. However, the EPI can assist charge trapping on shallow defects[43] created by non-equilibrium PLD growth or $V_O$s, in addition to the deep level trapping.

The polaronic reduction of $μ_{2DES}$ fundamentally limits the application potential of the STO-based heterostructures. As illustrated in Supplementary 2, this limit can possibly be circumvented through manipulation of $V_O$s. The $V_O$s inject into the 2DES extrinsic charge carriers[42] which increase the electron screening and thus reduce the EPI. This trend is consistent with the recent study at bare STO(100) surfaces[33] and, furthermore, explains the recent results on the $γ$-Al$_2$O$_3$/STO where the $V_O$s have not only increased $n_s$ but also $μ_{2DES}$ by almost two orders of magnitude. On the other hand, the $V_O$s can assist the EPI due to charge trapping on shallow defects[43] and also increase the defect scattering rate, both effects counteracting the above positive trend. Further experiments on oxygen-deficient LAO/STO will allow a better understanding of the role of $V_O$s and ways to optimize $μ_{2DES}$.

**Temperature dependence of polaronic effects**

An intriguing peculiarity of the charge carriers at the LAO/STO interface is the drop of $μ_{2DES}$ by more than one order of magnitude as the temperature increases above 200 K[18]. To unveil the underlying microscopic mechanism, we measured the temperature dependence of the $L_3$ angle-integrated EDC from Fig. 1e (the angle integration makes our analysis robust against extrinsic thermal scattering of high-energy photoelectrons, which averages the ARPES signal in $\mathbf{k}$-space[45]). These results are shown in Fig. 4a in comparison with experimental temperature dependence of $μ_{2DES}$ in Fig. 4b, characteristic of the LAO/STO samples[18], which was derived from Hall effect measurements. The experimental $A(\omega,\mathbf{k})$ in Fig. 4a shows that with increase of temperature, first, the polaronic hump broadens and propagates to higher binding energies. This trend is accompanied by increase of the QP peak width, Fig. 4d. These effects can be attributed to incoherent scattering on thermally populated phonon modes. A striking effect beyond this weight-conserving mechanism is, however, that the QP peak loses its integral weight $Z_0$ (Fig. 4c) and completely dissolves in the phonon hump of $A(\omega,\mathbf{k})$ towards 190 K (Fig. 4a). The decrease of $Z_0$, following the trend of the temperature dependent $\mu_{2DES}$, unambiguously signals that the QP spectral weight leaks to soft phonon modes, different from the hard LO3, with concomitant further increase of $m^*$. This phenomenon provides the microscopic mechanism of the puzzling $\mu_{2DES}$ drop with increasing temperature[18]. Interestingly, ARPES on the TiO$_2$(001) surface shows similar temperature effects[24] despite different crystallographic structure of TiO$_2$. In line with our results, an optical study of bulk STO[31] has also revealed reduction of the Drude weight with temperature. Surprisingly, the bare STO(100) surface at large $n_v$[32] does not show any systematic temperature effects in the QP weight; clearly, the LAO overlayer and 2DES significantly alter the EPI in the LAO/STO system.

Numerical analysis of the temperature dependent ARPES spectral shape allows us to identify the frequency $\omega_0"$ of the soft phonon mode dominating the EPI. From the normalized EDCs, Fig. 4a, we evaluate the temperature dependence of the spectral weight $Z_0(T)$ by Gaussian fitting of the QP peak. The resulting $Z_0(T)$ in Fig. 4c can then be fitted with the analytic formula for the independent boson model[1]

$$Z_0(T) = e^{-2g(2N+1)} I_0[2g(2N+1)], \quad (2)$$

where $N = \left(e^{\omega_0/T} - 1\right)^{-1}$ is the Bose filling factor, which describes the population number as a function of the mode frequency $\omega_0$ at $T$, and $I_0$ is the modified Bessel function. This equation, neglecting momentum dependence and thus valid for our momentum-integrated EDCs, describes transfer of the spectral weight from the QP peak to the hump with temperature. The constant $g$-factor, measuring the EPI strength, was set to 0.95 in order to reproduce the experimental value $Z_0(12K) \sim 0.4$. Fitting $Z_0(T)$ in the temperature range below ~120 K yields $\omega_0" \approx 18$ meV (solid line in Fig. 4c), while the high-$T$ range yields $\omega_0" \sim 14$ meV (dashed line in Fig. 4c). This crossover of $\omega_0"$ can be linked to the second-order tetragonal to cubic phase transition in STO at 105K[32,46]. One may therefore argue that the $\mu_{2DES}$ drop in LAO/STO has structural origin. Interestingly, the fastest change of the QP peak width in Fig. 4d falls into the same temperature region. All other possible temperature effects in the QP linewidth, including the Bloch-Grüneisen contribution due to acoustic phonons, are weaker for our system dominated by optical phonon, and obscured by the phase transition and limited energy resolution. We note that, strictly speaking, our $Z_0(T)$ model (2) had to include EPI with both low- and high-energy phonons, contributing to the total EPI. However, the result would not change significantly because the Bose filling factor $N$, determining the $Z_0(T)$ dependence, is dominated by the term $e^{\omega_0/T}$ most sensitive to small $\omega_0$. In other words, the $Z_0(T)$ dependence is most sensitive to the low-energy sector of the phonon spectrum.

Next we return to our DFT calculations to identify the observed soft $\omega_0''$ phonon mode. The phonon dispersions of the cubic (high-$T$) phase, Fig. 3a, show a multitude of low-energy modes. The lowest energy LO mode (LO1), which was previously linked with kinks in ARPES dispersions[28], has an energy of ~22 meV. This energy is considerably higher than our high-$T$ fitted $\omega_0'' \sim 14$ meV and, moreover, our calculations do not show a significant frequency difference between the cubic and tetragonal phases of STO. These facts, together with the reported small coupling constant $\lambda$[27], make the involvement of this LO1 mode unlikely. In Fig. 3a we see, however, the TO modes strongly affected by the electron doping $n_v$ due to enhanced screening. The TO1 mode, sketched in Fig. 3c, is a polar mode whose instability in undoped STO leads to its quantum-paraelectric behaviour[47,48]. Its frequency rapidly increases and becomes real with increase of $n_v$, stabilizing above our actual $n_v \sim 0.12$ electrons/u.c. at $\omega_{TO1} \sim 13.7$ meV at the $\Gamma$ point. This $\omega_{TO1}$ matches well with our $\omega_0'' \sim 14$ meV fitted in the high-$T$ range. Turning to the low-$T$ range, previous calculations on undoped STO have shown that the tetragonal phase transition increases $\omega_{TO1}$[49]. Our computations with $n_v = 0.12$ electrons/u.c. for the tetragonal phase indeed show $\omega_{TO1}$ to shift to 15.3 meV for the doubly degenerate mode and 18.1 meV for the non-degenerate mode along the octahedral rotation axis, which is in good agreement with our low-$T$ fitted value $\omega_0'' \sim 18$ meV. Octahedral rotations different from this bulk tetragonal distortion were observed at the LAO/STO interface due to the electrostatic doping[34,35], but since they are combinations of distortions similar to the one studied here, they are expected to also lead to an upwards shift of the TO1 frequency. Based on the $t_{1u}$ symmetry of the TO1 mode, its good agreement with the experimental $\omega_0''$ and its sensitivity to the cubic to tetragonal phase transition, we associate the experimental soft phonon with TO1. Moreover, significant polaronic coupling to this polar mode is consistent with gigantic dielectric constant $\varepsilon_0$ of STO caused by large polar ionic displacements under electric field in this material on the verge of ferroelectric instability[30]. The TO1 mode is associated with *long-range EPI*, as evidenced by the increase of $\frac{\partial \omega_0''}{\partial n_v}$ towards the $\Gamma$-point. The hard LO3 and soft TO1 modes involved in the polaron formation are therefore associated with opposite types of EPI. The EPI strength in the latter case can be estimated from the Fröhlich model which, although strictly valid only for the polar LO modes, represents the only viable model to find the long-range coupling constant $\alpha$. Based on exact diagrammatic QMC calculations[50] and $Z_0$ at the top of our temperature range, we estimate $\alpha \approx 2$ for the TO1 mode. Similarly to the above LO3 phonon, this value stays below the weak- to strong-coupling crossover at $\alpha \approx 6$, which implies that the TO1 phonon is also consistent with the large polaron scenario. We note that if LAO/STO superconductivity is driven by a phonon mechanism[51], it can be related to the discovered polaronic activity. Whereas the hard LO3 phonon energy much exceeds the energy scale of the superconducting transition at 0.3K, involved in the electron pairing may be the soft TO1 phonon.

The 2DES at the LAO/STO interface realizes thus a polaronic metal state involving at least two phonons with different energies and thermal activity. The hard LO3 phonon at $\omega_0' \sim 118$ meV, associated with short-range EPI, is directly resolved as the characteristic hump in the experimental $A(\omega,\mathbf{k})$. It sets the fundamental limit of $\mu_{2DES}$ at low temperature, and exhaustively accounts for the $m^*$ renormalization without notable effects of electron correlations. Another soft phonon, likely the TO1 one associated with long-range EPI, changes its frequency from $\omega_0'' \sim 18$ to 14 meV across the phase transition in STO. This phonon causes a dramatic fading of the QP weight with temperature as the microscopic mechanism behind the $\mu_{2DES}$ drop observed in transport. The two phonons form a large polaron characterized by a lattice deformation extending over several unit cells. The polaronic activity is typical of oxide perovskites, reflecting their highly ionic character and easy structural transformations[30,41]. Our discovery may have implications for other related oxide systems, including LAO/STO interfaces with different crystallographic orientations. In a methodological perspective, we have demonstrated the power of the newly emerging experimental technique of ultrahigh-resolution SX-ARPES to retrieve information about polaronic effects at buried interfaces in the most direct way as embedded in one-electron $A(\omega,\mathbf{k})$.

**Methods**

Our LAO/STO samples were grown using PLD (for details of the growth procedure see Refs. 19,22) and subsequently annealed in oxygen atmosphere with a pressure of 200 mBar at 550°C for one hour to minimize the concentration of $V_{OS}$. These vacancies manifest themselves in ARPES spectra as characteristic dispersionless in-gap states at a binding energy of ~1.2 eV[14,22] which, similarly to bare STO, grow with exposure to X-rays. Our resonant spectra in Fig.1a show only traces of such spectral structures enhanced at the $Ti^{3+}$ $L_3$- and $L_2$-resonances, indicating negligible concentration of $V_{OS}$. Measurements on samples with various $n_s$ varied through the $V_{OS}$ (Supplementary 2) demonstrated constant energy of the 118-meV spectral peak, excluding its plasmonic origin. The temperature dependent mobility data were derived from standard Hall effect and conductivity measurements performed in the van der Paw geometry on square samples.

SX-ARPES experiments were performed at the ADRESS beamline of the Swiss Light Source, Paul Scherrer Institute, Switzerland[52]. The experimental geometry allows symmetry analysis of the valence states using variable linear polarizations of incident X-rays. The experiments are normally performed at low sample temperatures around 12K to quench thermal scattering of high-energy photoelectron destructive for the coherent **k**-resolved spectral component[45]. The combined (beamline and analyzer) energy resolution was set to 80 meV for measurements of the FS, and to 40 meV for high-resolution measurements of the band dispersions. Such resolution achieved for an interface buried behind a ~18 Å thick overlayer presents the forefront of nowadays SX-ARPES instrumentation. The temperature dependence was measured with increasing temperature in order to avoid possible hysteresis effects[53].

Band structure calculations were performed using the pseudo-self-interaction correction (pSIC) method[54], with a plane wave basis set and ultrasoft pseudopotentials. This ab-initio approach corrects the main deficiencies of basic density functional theory for a vast range of oxides[8]. The theoretical band structure and the FS were calculated for an interfacial carrier density of 0.115 electrons/unit cell. This value is consistent with those determined using Hall effect measurements for the LAO/STO interface. The corresponding Fermi momentum $k_F$ of the $d_{yz}$-states is in good agreement with the experimental value of ~0.29 Å$^{-1}$ determined as the highest band dispersion point.

DFT-based phonon calculations under different electron doping were performed using the VASP code[56]. The electron count was adjusted while adding a compensating background charge. We used the PBEsol functional, which gives reliable lattice parameters and phonon frequencies[55], and PAW potentials[56] with Sr($4s,4p,5s$), Ti($3p,3d,4s$) and O($2s,2p$) valence shells. Phonons were then computed using the frozen phonon approach[57]. For further details, see Ref. 49.

**Acknowledgements**

We thank P.R. Willmott, J. H. Dil, R. Claessen, J.-M. Triscone, F. Baumberger, F. Bisti, Z. Wang, M. Radović, J. Mesot and N.A. Spaldin for fruitful discussions, and X. Wang for software support of the experiment. Parts of this research were supported by the ImPACT Program of the Council for Science, Technology and Innovation (Cabinet Office, Government of Japan), and the ETH Zürich and ERC Advanced Grant program, No. 291151.

**Author contributions**

C.C. and V.N.S. have performed the experiment supported by V.A.R. and T.S. and processed the data. C.C. has grown the samples. V.N.S. and A.S.M. have developed the scientific concept. A.S.M. has performed the polaronic analysis supported by O.S.B. and N.N. A.F. and U.A. have calculated the electron and phonon band structures, respectively. All authors have discussed the results and interpretations as well as the manuscript written by V.N.S. and A.S.M. with contributions of C.F., U.A. and C.C.

**Figure captions**

**Fig. 1. Experimental low-temperature (12K) overview SX-ARPES data. a**, Resonance (angle-integrated) photoemission intensity map, identifying the 2DES signal at the $L_3$ and $L_2$ resonances of the interface Ti ions. **b**, FS map at the $L_3$-resonance, where the superimposed theoretical FS contours identify the $d_{xy}$ (pink) and $d_{yz}$ (green) sheets.

**Fig. 2**. **Experimental low-temperature (12K) high-resolution SX-ARPES. a,b**, High-resolution ARPES images along the ΓX ($k_y$=0) line at the $L_3$- and $L_2$-edges, respectively, with the superimposed theoretical $d_{xy}$ (pink) and $d_{yz}$ (green) bands. The lower panels show the corresponding second derivative -d$^2I$/d$E^2$>0 plots, which clearly show both the quasi-particle (QP) peak and the dispersive hump formed by the LO3 phonon. **c,d** A series of (normalized) EDCs extracted from **a,b**, respectively, at the indicated $k_x$-values through the occupied part of the BZ. The two curves at the bottom show EDCs integrated over the **k**-ranges indicated in **c,d** as well as the whole BZ in the $k_x$-direction. The characteristic PDH spectral structure in **c,d** manifests a polaronic metal state formed by the hard LO3 phonon and renormalizing the $d_{yz}$-band dispersion in **a,b**, and the clear hump dispersion in **b** identifies a large polaron.

**Fig. 3. Theoretical phonon modes in cubic-phase doped STO. a,** Phonon dispersion at various electron doping levels $n_v$, with our case corresponding to $n_v \sim 0.12$. The arrows indicate the TO1 and TO2 modes shifting as a function of $n_v$. The imaginary modes at the R and M-points represent different octahedral rotation instabilities, whereas the one at the Γ-point in the undoped materials is the polar (quantum-paraelectric) instability. **b,c,** Atomic displacements associated with the breathing LO3 mode at the R-point and the polar TO1 mode at the Γ-point, respectively.

**Fig. 4. Temperature dependence of polaronic effects. a**, Angle-integrated EDCs at the $L_3$-resonance acquired for temperatures between 12 and 190 K. The QP peak dissolving into the hump towards ~190K explains **b**, the mobility drop with temperature measured in transport (the dashed line trends through the experimental points). **c**, Temperature dependent QP spectral weight $Z_0(T)$ fitted by the independent boson model of Eq. (2). The fit identifies a soft phonon mode (likely the TO1) with $\omega_0'' = 18$ meV (solid line) in the low-$T$ region, which shifts to $\omega_0'' = 14$ meV (dashed) in the high-$T$ region through the tetragonal to cubic phase transition in STO. The fading colours represent the validity ranges of the fits. **d**, QP peak width (including the instrumental resolution).

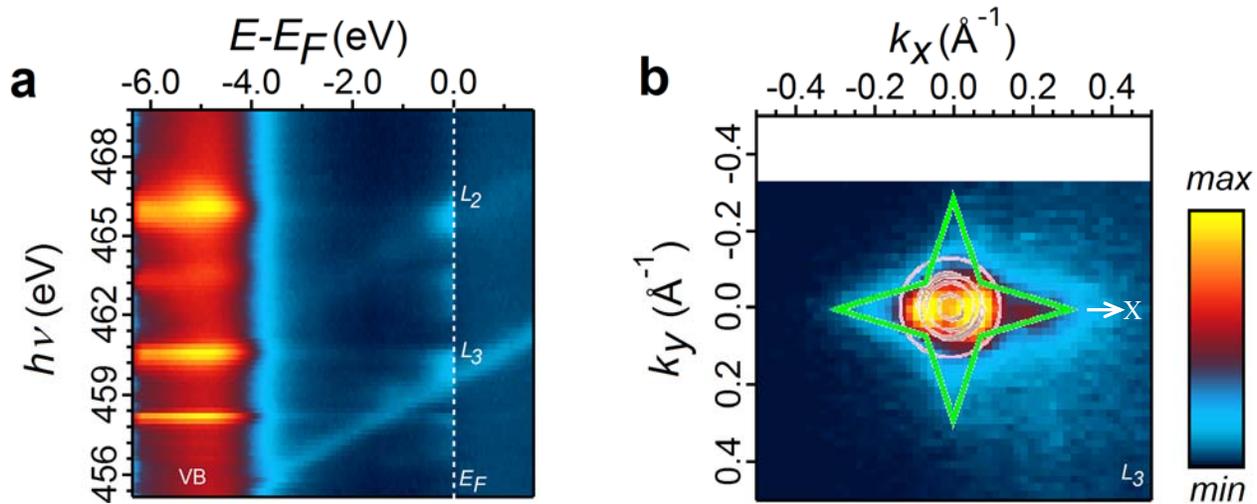

Fig. 1

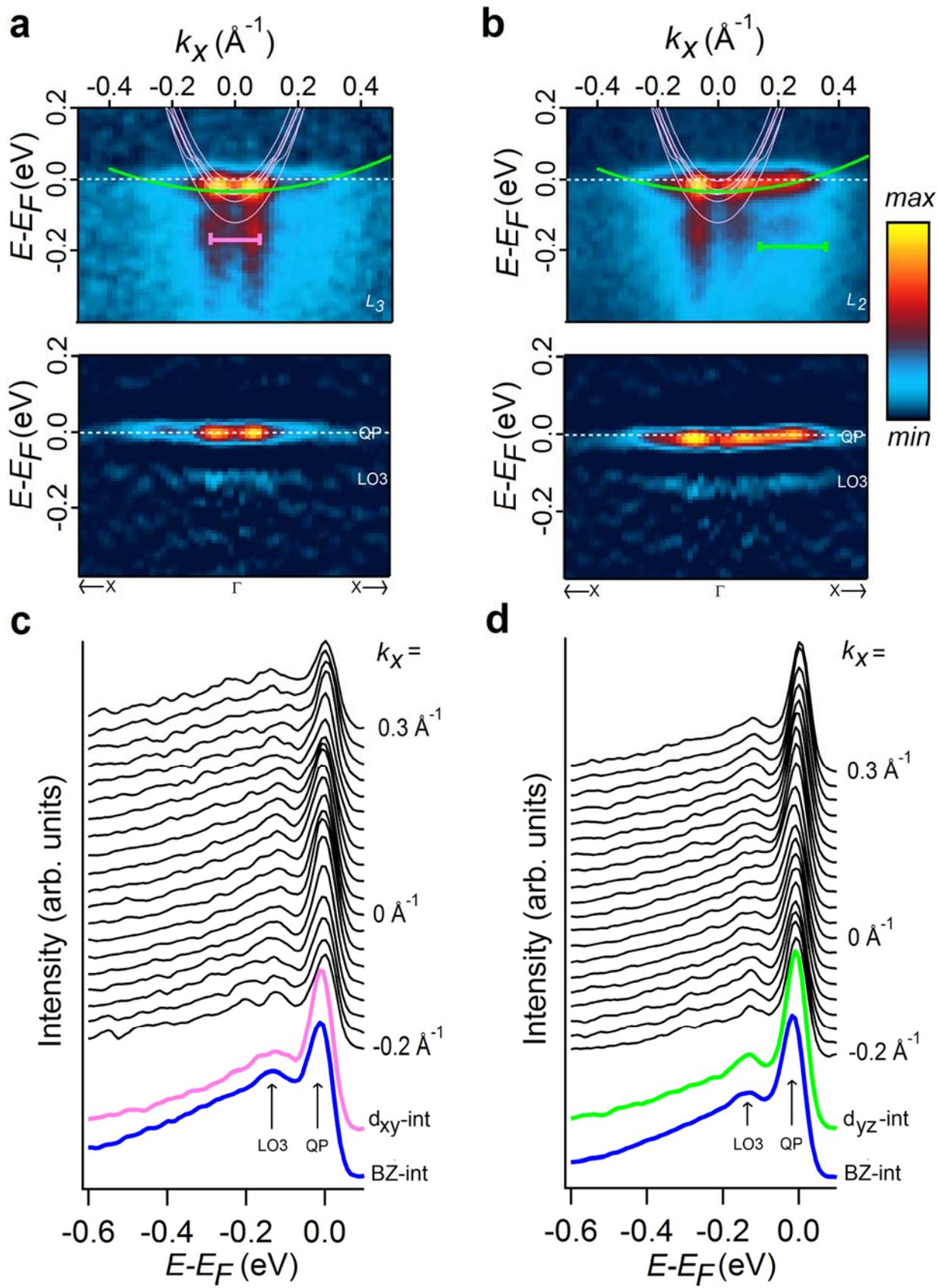

Fig. 2

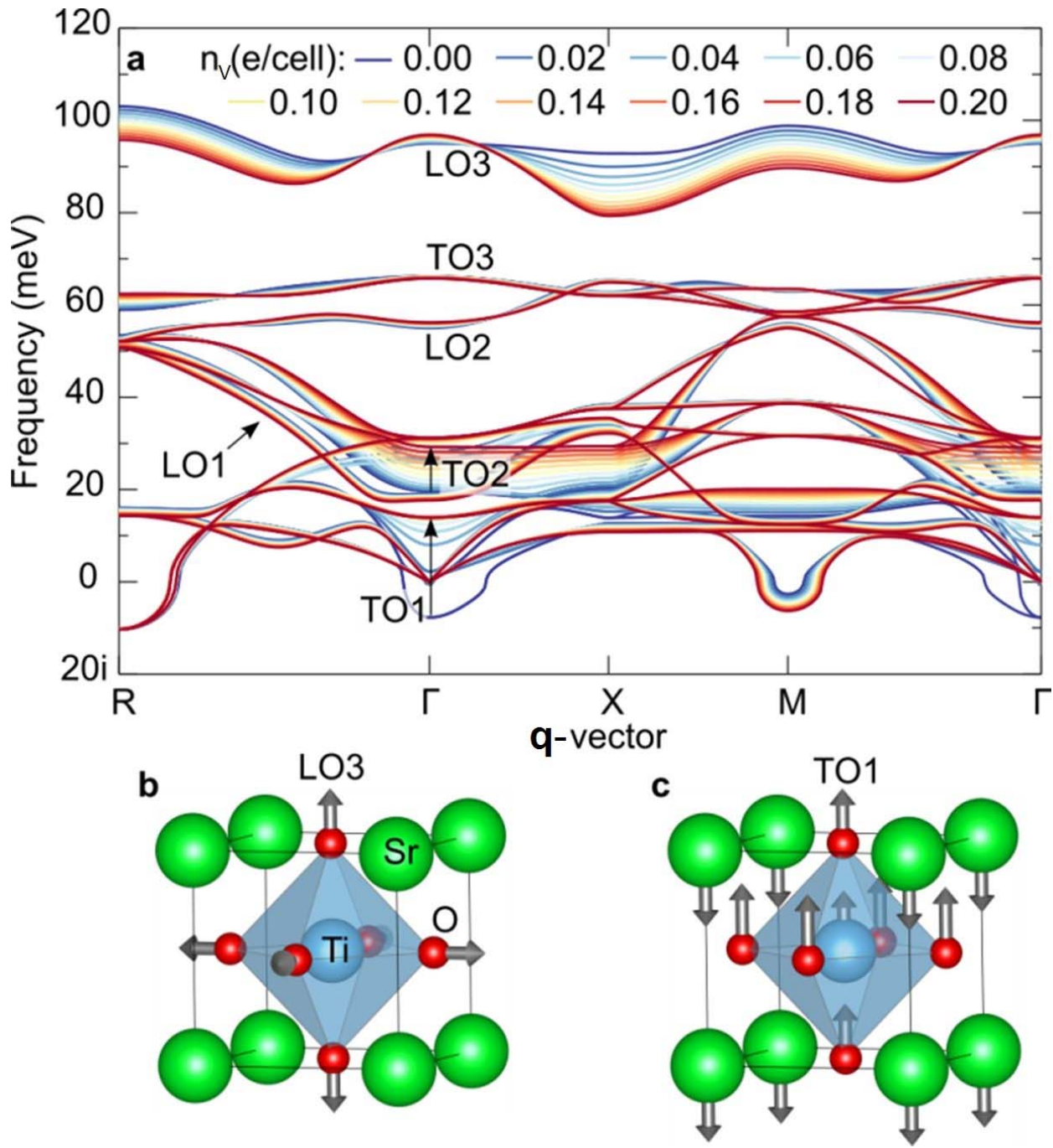

Fig. 3

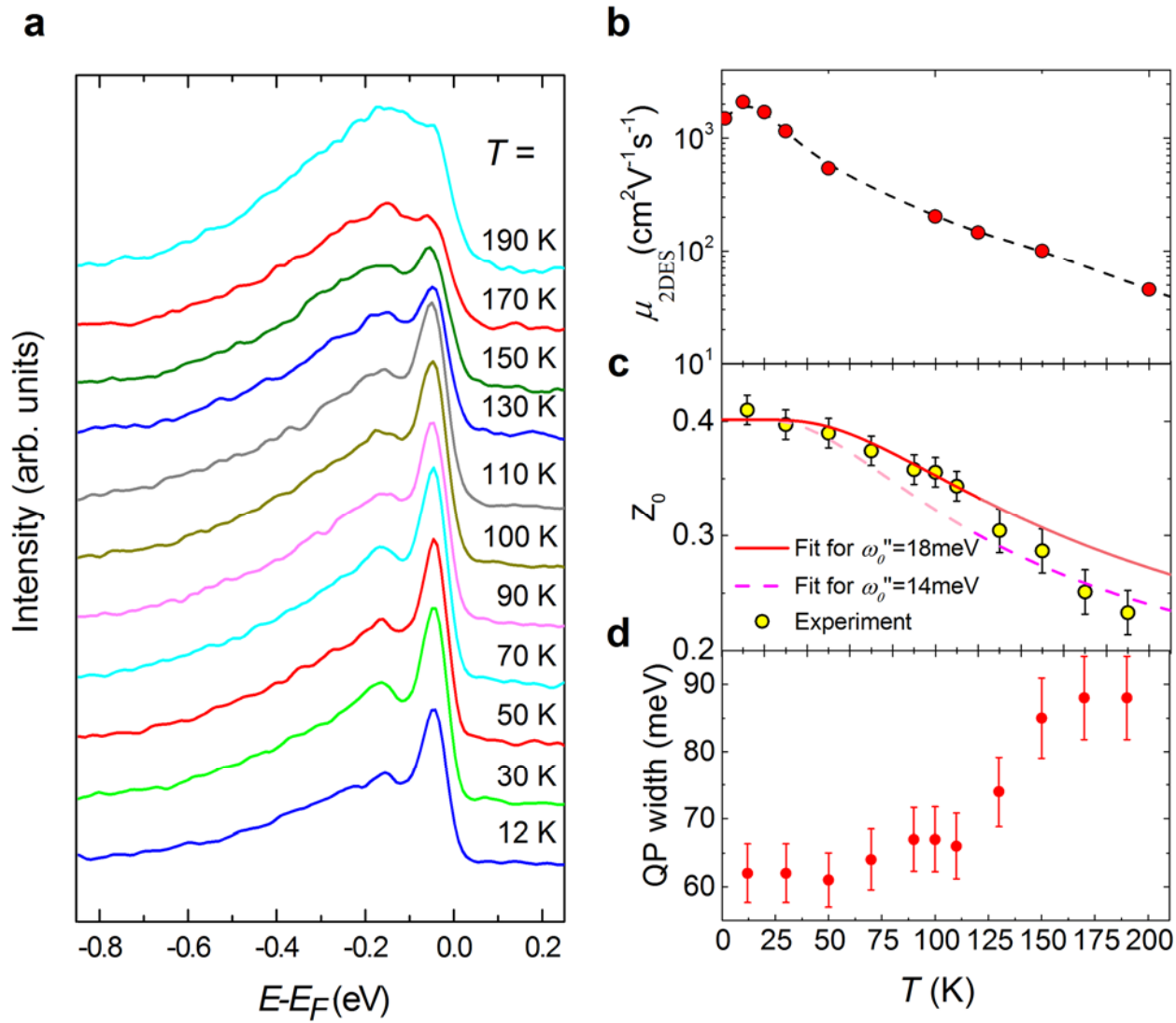

Fig. 4

# Supplementary 1: SX-ARPES data for *p*-polarized X-rays

Here, we present our SX-ARPES data acquired at 12K with *p*-polarized X-rays, parallel to the *s*-polarization data reported in the main text, Figs. 1 and 2. We select now the $d_{xz}$-derived states, symmetric relative to the ΓX line of the two-dimensional BZ. The resonance map of (angle-integrated) photoemission intensity, Fig. S1a, again identifies the 2DES signal at $E_F$ blowing up near the two $Ti^{3+}$ $L_3$- and $L_2$-resonances, although its intensity is smaller compared to the *s*-polarization because of the missing strong $d_{xy}$-intensity. The FS map in Fig. S1b acquired at the stronger $L_3$-resonance clearly displays the elliptical $d_{xz}$-sheet extending in the $k_y$ –direction and derived from the $d_{xz}$-derived state symmetric relative to the ΓX line. Consistently with this map, the ARPES images measured along the ΓX line at the $L_3$- and $L_2$-resonances, Fig. S1c and d respectively, display the $d_{xz}$-derived band with its smaller $k_F$ along the ΓX line compared to the $d_{yz}$-state in Fig. 1. In the $L_2$-image we note remnant intensity from the antisymmetric $d_{yz}$-derived band, which creeps in due to slight relaxation of the symmetry selection rules caused by the tetragonal distortion of STO at low temperatures. Importantly, the ARPES images and the corresponding EDCs in Fig. S1e and f again reveal the pronounced PDH structure of $A(\omega,\mathbf{k})$ with the LO3-related polaronic hump at ~118 meV.

**Fig. S1. Experimental low-temperature (12K) SX-ARPES results collected with *p*-polarization. a**, Resonance photoemission intensity map, identifying the 2DES signal at the $L_3$ and $L_2$ resonances. **b**, FS map at the $L_3$-resonance, showing the $d_{xz}$-derived sheet. **c,d**, High-resolution ARPES images along the ΓX line at the $L_3$- and $L_2$-edges, showing the $d_{xz}$-derived band. **e,f** A series of (normalized) EDCs extracted from **c,d**, respectively, at the indicated $k_x$-values. The two curves at the bottom show EDCs integrated over the whole BZ in the $k_x$-direction. The *p*-polarization data confirms the characteristic PDH spectral structure manifesting a polaronic metal state formed by the hard LO3 phonon.

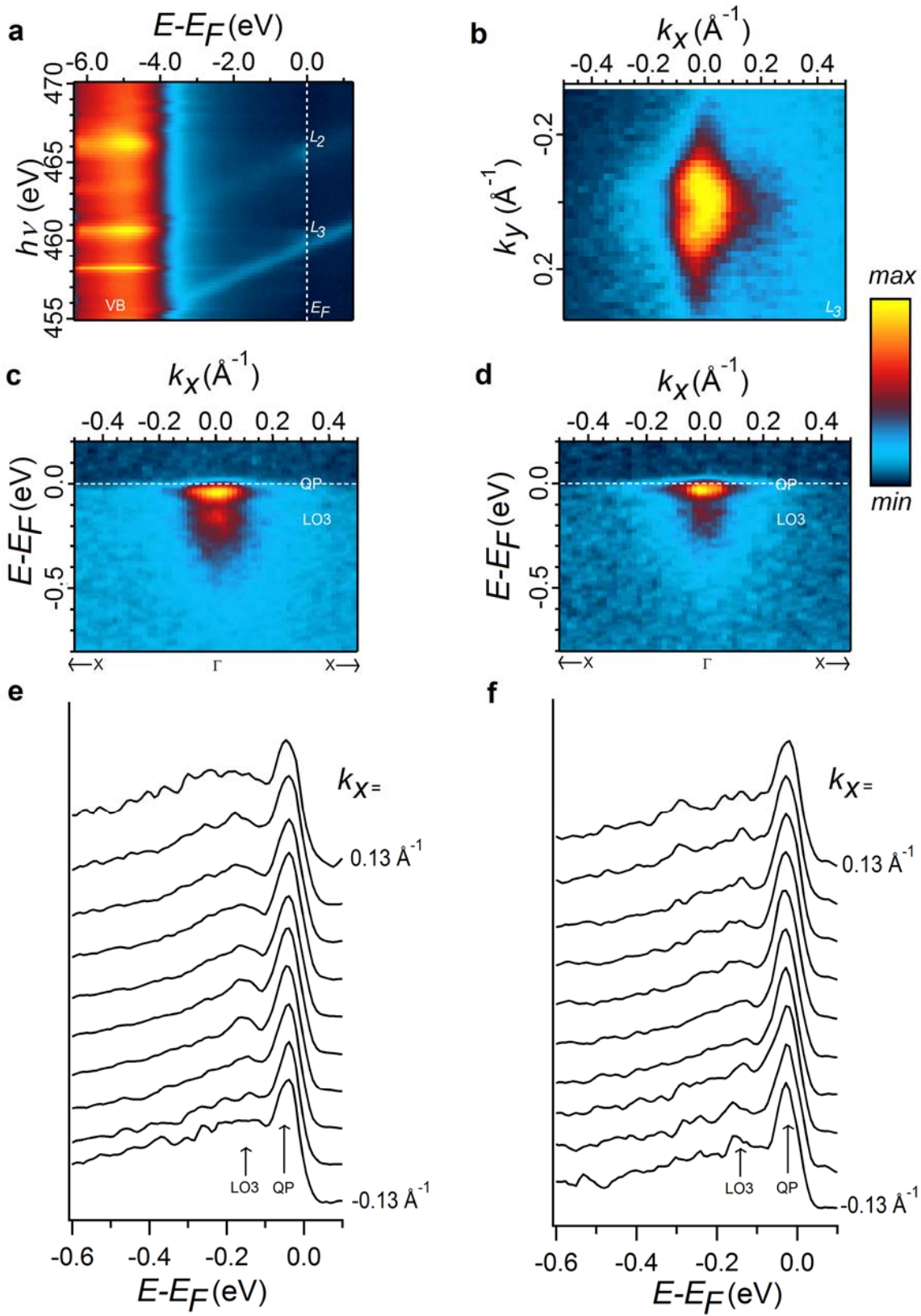

# Supplementary 2: SX-ARPES data for oxygen-deficient LAO/STO samples

Here, we present SX-ARPES data on oxygen-deficient (OxD) LAO/STO samples to confirm the polaronic origin of the peak-dip-hump spectral structure, and illustrate a possibility to manipulate the EPI through the $V_O$s. Post-annealing in high pressure of oxygen is an important step in the fabrication of standard LAO/STO samples, which reduces the number of $V_O$s produced during the non-equilibrium PLD growth[19]. The simplest way to obtain OxD samples is therefore to exclude the last post-annealing step. The main effect of the $V_O$s in STO is to inject extrinsic charge carriers and thus increase $n_s$ of the interface 2DES (see, for example, Ref. 42).

We have prepared OxD samples using the same growth conditions as described in the Methods section of the main text, but instead of the post-annealing in oxygen the samples were simply cooled down at room temperature at the same oxygen pressure $10^{-4}$ mbar as used during the deposition. Fig. S2 shows the SX-ARPES experimental results collected on these samples under the same conditions (experimental geometry, X-ray polarization, resolution and low sample temperature) as the standard oxygen-annealed samples in Fig. 2. The experimental ARPES image in Fig. S2a demonstrates the expected increase of $k_F$ compared to the standard sample, and fading of the waterfall intensity. Furthermore, the corresponding $k_x$-integrated spectrum in Fig. S2b reveals scaling up of broad spectral intensity around -1.2 eV, the spectroscopic signature of the in-gap states induced by the $V_O$s[22,23], accompanied by dramatic reduction of the polaronic hump. We note that the OxD samples are highly sensitive to X-ray irradiation, acting to multiply $V_O$s as evidenced by increase of the in-gap spectral intensity and reduction of the polaronic hump with exposure time. The present data have been collected under saturation, achieved after about 15 min of X-ray irradiation with a photon flux of $\sim 10^{13}$ photons/sec/0.01% bandwidth[S1]

Our spectroscopic comparison of the standard and OxD samples, first, identifies the bosonic mode forming the peak-dip-hump spectral structure. To retrieve the hump energy for the OxD sample, we have fitted the experimental $k_x$-integrated EDC in Fig. S2b by three Gaussians, representing the in-gap states, hump and the QP peak (whose lineshape is anyway limited by the experimental resolution Gaussian). The fit returns the hump energy 100±20 meV relative to the QP, which is within the error bars identical to the 118 meV for the standard sample. If the hump had the plasmon origin, its energy would scale proportional to $\sqrt{n_s}$. With the $n_s$ values $2.5 \cdot 10^{14}$ and $7.5 \cdot 10^{13}$ e/cm² determined from the experimental $k_F$ values (see Methods in the main text) for the OxD and standard samples, respectively, the plasmon frequency would change by a factor ~1.7 which is obviously not the case. Therefore, the constant energy of the hump rules out its plasmonic origin. The same conclusion has been made in ARPES experiments on bare STO(100) surfaces[33]

Second, our results suggest a possibility to circumvent the polaronic limit of $\mu_{2DES}$ through manipulation of OVs. As explained in the main text, the $V_O$s inject into the 2DES extrinsic charge carriers[42] which increase the electron screening and thus reduce the EPI strength. This effect explains, for example, the recent results on the $\gamma$-Al$_2$O$_3$/STO where the $V_O$s have increased $\mu_{2DES}$ by almost two orders of magnitude. On the other hand, the $V_O$s can assist the EPI due to charge trapping on shallow defects[43] and also increase the defect scattering rate, both effects counteracting the above positive trend. We note that the effect of the $V_O$s is actually beyond the simple doping picture restricted to changing of the band filling within the rigid band shift model. In particular, the $V_O$s increase the spatial extension of the $d_{xz/yz}$ bands into the STO bulk from ~50 Å for oxygen-annealed samples to ~150 Å and more[8,9,S2,S3] resulting in predominantly bulk conductivity and loss of the 2D nature of the LAO/STO interface system. Furthermore, the manipulation of the $V_O$s is complicated by diffusion processes which are hard to precisely control. Further experiments on OxD-LAO/STO interfaces will allow a better understanding of the role of $V_O$s and ways to optimize $\mu_{2DES}$.

**Additional references**

S1. V.N. Strocov, T. Schmitt, U. Flechsig, T. Schmidt, A. Imhof, Q. Chen, J. Raabe, R. Betemps, D. Zimoch, J. Krempasky, X. Wang, M. Grioni, A. Piazzalunga and L. Patthey. High-resolution soft X-ray beamline ADRESS at the Swiss Light Source for resonant inelastic X-ray scattering and angle-resolved photoelectron spectroscopies. *J. Synchr. Rad.* **17**, 631 (2010)
S2. G. Herranz, M. Basletić, M. Bibes, C. Carrétéro, E. Tafra, E. Jacquet, K. Bouzehouane, C. Deranlot, A. Hamzić, J.-M. Broto, A. Barthélémy and A. Fert. High Mobility in LaAlO$_3$/SrTiO3 heterostructures: Origin, dimensionality, and perspectives. *Phys. Rev. Lett.* **98**, 216803 (2007)
S3. M. Basletic, J.-L. Maurice, C. Carrétéro, G. Herranz, O. Copie, M. Bibes, É. Jacquet, K. Bouzehouane, S. Fusil and A. Barthélémy, Mapping the spatial distribution of charge carriers in LaAlO$_3$/SrTiO$_3$ heterostructures. *Nature Mat.* **7**, 621 (2008)

**Fig. S2. Experimental SX-ARPES results on OxD-LAO/STO samples. a**, High-resolution ARPES image along the $\Gamma X$ line at the $L_3$-edge, showing larger $k_F$ and thus $n_s$. **b** The corresponding $k_x$-integrated EDC (*bottom*) compared with that of the standard sample (*top*). Constant energy of the hump proves its polaronic origin, and its scaling down illustrates a possibility to manipulate the EPI through the $V_O$s.

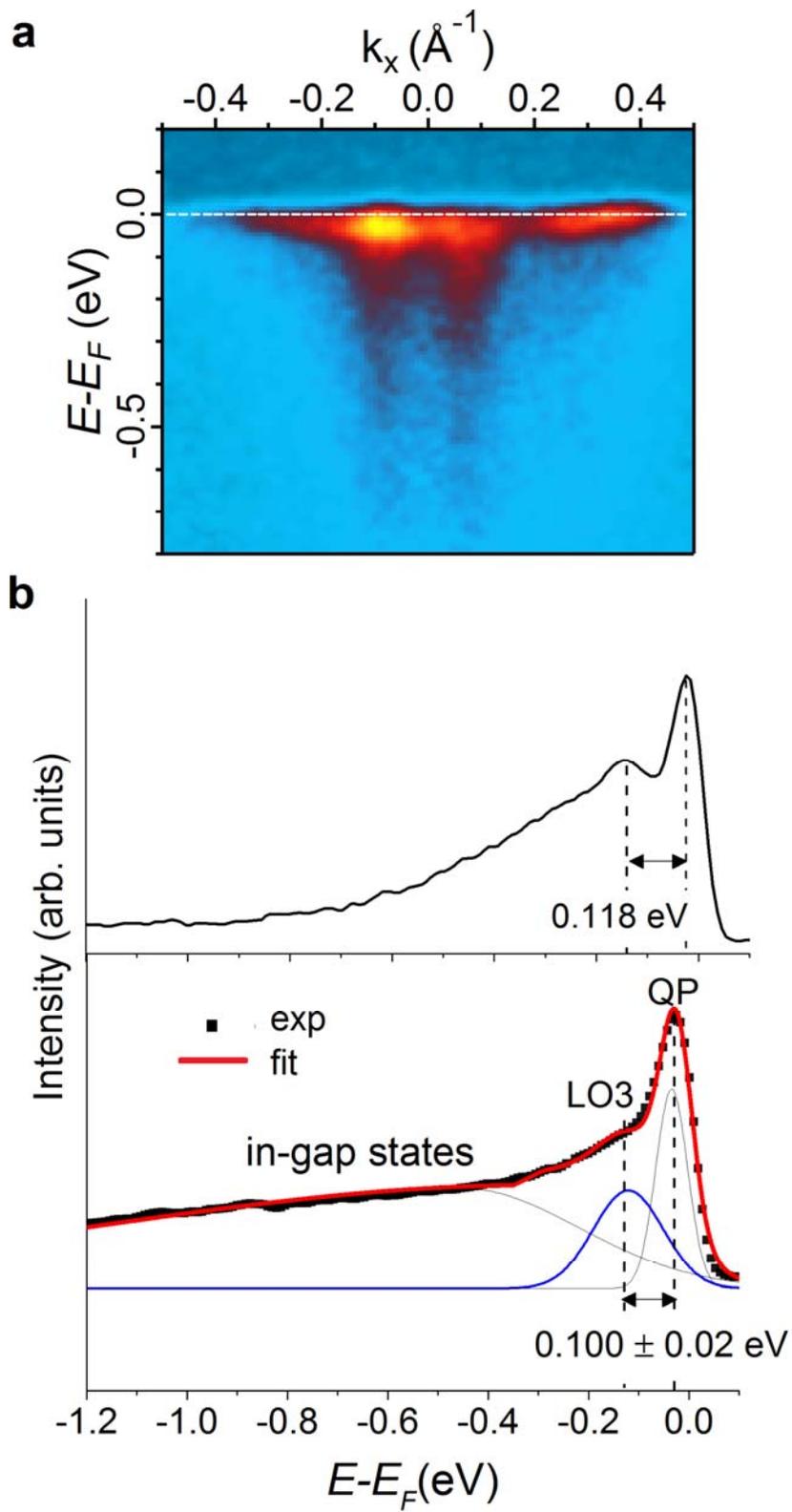

Fig. S2